# Adaptive Multi-Step Prediction based EKF to Power System Dynamic State Estimation

Shahrokh Akhlaghi, *Student Member, IEEE,* Ning Zhou, *Senior Member, IEEE,*

*Abstract*— Power system dynamic state estimation is essential to monitoring and controlling power system stability. Kalman filtering approaches are predominant in estimation of synchronous machine dynamic states (i.e. rotor angle and rotor speed). This paper proposes an adaptive multi-step prediction (AMSP) approach to improve the extended Kalman filter's (EKF) performance in estimating the dynamic states of a synchronous machine. The proposed approach consists of three major steps. First, two indexes are defined to quantify the nonlinearity levels of the state transition function and measurement function, respectively. Second, based on the nonlinearity indexes, a multi prediction factor ($M_p$) is defined to determine the number of prediction steps. And finally, to mitigate the nonlinearity impact on dynamic state estimation (DSE) accuracy, the prediction step repeats a few time based on $M_p$ before performing the correction step. The two-area four-machine system is used to evaluate the effectiveness of the proposed AMSP approach. It is shown through the Monte-Carlo method that a good trade-off between estimation accuracy and computational time can be achieved effectively through the proposed AMSP approach.

*Index Terms*--Dynamic state estimation (DSE), extended Kalman filter (EKF), phasor measurement unit (PMU), power system dynamics, non-linearity test.

## I. INTRODUCTION

Accurate estimation of power system dynamic state (rotor angle and speed of a synchronous machine) is important for monitoring and controlling transient stability. Dynamic states estimation (DSE) can provide prediction for generators, loads and controller. These states completely determine the behavior of the power system according to differential algebraic equations. Power system dynamic state are used to design the power system stabilizers and voltage regulators to maintain and improve transient stability. Increasing demand for reliable and sustainable energy pushes further the investigation of new techniques to improve the accuracy of the DSE for reliable and stable operation of a power system [1]-[2].

With the deployment of the phasor measurement units (PMU), because of high sampling rates and well synchronized data, the more and more studies on DSE have been carried out [3]-[4]. Based on literature, in most of the approaches dealing with DSE using PMU data, Kalman filtering (KF) techniques are dominant because of their high computational efficiency [5]–[17]. For instance, Huang et al. [5] studied the feasibility of applying an extended Kalman filtering (EKF) to estimate the dynamic states using PMU data. Ghahremani and Innocent [6] proposed a decentralized EKF approach to estimate the dynamic states and unknown inputs simultaneously. Zhao et al. [7] developed an iterated EKF based on the generalized maximum likelihood approach to estimate the dynamic states during disturbances. Akhlaghi et al. [9]-[10] proposed an adaptive interpolation approach to mitigate the impact of non-linearity on DSE. The approach adaptively adjusted interpolation factors to strike a balance between estimation accuracy and computation time. [11]-[15] proposed the unscented Kalman filtering to estimate power system dynamic states. Zhou et al. [16] proposed an ensemble Kalman filter approach to simultaneously estimate the dynamic states and parameters. These studies have laid a solid ground for estimating the dynamic states of a power system. Among these methods, the EKF is often considered a standard benchmark for DSE because of its simplicity and high computational efficiency [17].

Studies in [10] and [18] revealed that the severe non-linearity could reduce the effectiveness of the EKF in DSE. This is because the EKF approximates a non-linear function using 1$^{st}$ order Taylor expansion and eliminate the higher order. "When a system is linear and has Gaussian noises, the EKF is equivalent to the standard KF and can produce optimal estimates. However, when the non-linearity of a system is high, the EKF shows very inaccurate estimation" [18]. According to the studies carried out by [9] and [10], in the PMUs typical sampling rates (30 and 60 sample/s), the estimation accuracy of the EKF can degrade. To improve the EKF's performance in case of a system with severe non-linearity, [18]-[21] proposed to increase effective sampling rates by adding pseudo-measurements through linear interpolation. [22] proposed to decrease sampling intervals by a multi-step prediction approach. It was shown that linear interpolation and multi-step prediction can significantly improve the estimation accuracy.

The classical EKF consists of two steps: predication and correction. The multi-step prediction approach repeats the prediction step for a few time cycles before the correction step. To apply the multi-step approach, an important factor is multi predication factor '$M_p$', which is the number of times that the predication step is repeated before the correction step. On one hand, more prediction steps can reduce the negative impact of the non-linearity on estimation accuracy. On the other hand, more prediction steps demand additional calculation which incurs heavier computation burden. To achieve a good trade-off between computation time and estimation accuracy, '$M_p$' shall be large enough to mitigate the negative impact of the nonlinearity on estimation accuracy and small enough to avoid adding excessive computational burden.

In [22], '$M_p$' was predetermined as a constant before the execution of the EKF. This practice is acceptable when the system's non-linearity does not change significantly. Yet, in a power grid, the non-linearity often changes with changing operation points and conditions. When non-linearity levels change, a constant '$M_p$' may not always be able to achieve a good trade-off between estimation accuracy and computation time. To avoid such drawbacks, this paper proposes an *adaptive multi-step predication* (**AMSP**) approach, which adaptively adjusts '$M_p$' based on non-linearity levels. Specifically, under conditions with severe non-linearity, '$M_p$' shall be larger to improve estimation accuracy. Under linear or quasi-linear conditions, '$M_p$' can be smaller to reduce computation time and still maintain the same level of accuracy. The proposed **AMSP** approach consists of three steps. First, two indexes are determined to quantify the non-linearity levels of the state transition function and measurement function. Second, '$M_p$' is determined according to the proposed non-linearity indexes. Finally, '$M_p$' number of prediction steps are repeated before performing the correction step. The objective is to make a well-informed trade-off between estimation accuracy and computation time.

The rest of paper is organized as follows. Section II reviews the dynamic models used for estimation. The proposed non-linearity indexes are presented in Section III. Section IV proposes the **AMSP** approach and applies it to EKF. Section V discusses the study approaches. Case studies are presented in Section V. Conclusions are drawn in Section VI.

## II. DYNAMIC STATE ESTIMATION MODEL

This section introduces the dynamic model of a synchronous machine to be used by the proposed **AMSP** approach for DSE of a synchronous machine. The 4[th] order differential equations in a local *d-q* reference frame is used as follows (readers may refer to [10] and [17] for more details):

$$\begin{cases} \dfrac{d\delta}{dt} = \omega_0 \Delta\omega_r & \text{(1.a)} \\ \dfrac{d\Delta\omega_r}{dt} = \dfrac{1}{2H}(T_m - T_e - K_D \Delta\omega_r) & \text{(1.b)} \\ \dfrac{de'_q}{dt} = \dfrac{1}{T'_{d0}}\left(E_{fd} - e'_q - (x_d - x'_d)i_d\right) & \text{(1.c)} \\ \dfrac{de'_d}{dt} = \dfrac{1}{T'_{q0}}\left(-e'_d + (x_q - x'_q)i_q\right) & \text{(1.d)} \end{cases}$$

In (1), the states $\delta$, $\Delta\omega_r$, $e'_d$ and $e'_q$ are the rotor angles, rotor speeds and transient voltages along *d* and *q* axes, respectively. $\omega_0 = 2\pi f_0$ is the synchronous speed; parameters $H$ and $K_D$ are the inertia and damping factor, respectively; $T_m$ and $T_e$ are the mechanical and the electric air-gap torque in per-unit (*pu*); and $E_{fd}$ is the internal field voltage. Variables $x_d$ and $x_q$ are the synchronous reactance; $x'_d$ and $x'_q$ are the transient reactance, which all are along *d* and *q* axes. $i_d$ and $i_q$ are the stator currents along *d* and *q* axes. $T'_{d0}$ and $T'_{q0}$ are the open circuit time constants in the *dq0* frame.

To facilitate the notation of applying the EKF to power system DSE, a modified Euler method is applied to (1) to obtain a discrete model as shown in (2) and (3), with sampling interval of $\Delta t$ [17]. In (3) the subscript *k* is the time index, which indicates the time instance at $k\Delta t$.

$$\begin{cases} x_k = \Phi(x_{k-1}, u_{k-1}) + w_{k-1} & \text{(2.a)} \\ z_k = h(x_k, u_k) + v_k & \text{(2.b)} \end{cases}$$

$$x_k = \begin{bmatrix} \delta & \Delta\omega & e'_q & e'_d \end{bmatrix}^T \quad \text{(3.a)}$$

$$u_k = \begin{bmatrix} T_m & E_{fd} & i_R & i_I \end{bmatrix}^T \quad \text{(3.b)}$$

$$z_k = \begin{bmatrix} e_R & e_I \end{bmatrix}^T \quad \text{(3.c)}$$

Here $x_k$, $u_k$, and $z_k$ are the state, known input and measurement vectors, respectively. Functions $\Phi(*)$ and $h(*)$ are the state transition function and the measurement function, respectively. To transform (1) into general state space model (2) and implement the state transition function and measurement function, $i_R$, $i_I$, $e_R$ and $e_I$ were written as functions of *x* and *u* using (4) and (5).

$$\begin{bmatrix} i_d \\ i_q \end{bmatrix} = \begin{bmatrix} \sin\delta & -\cos\delta \\ \cos\delta & \sin\delta \end{bmatrix} \begin{bmatrix} i_R \\ i_I \end{bmatrix} \quad \text{(4)}$$

$$\begin{bmatrix} e_R \\ e_I \end{bmatrix} = \begin{bmatrix} \sin\delta & \cos\delta \\ -\cos\delta & \sin\delta \end{bmatrix} \begin{bmatrix} e'_d \\ e'_q \end{bmatrix} - \begin{bmatrix} \cos\delta & -\sin\delta \\ \sin\delta & \cos\delta \end{bmatrix} \begin{bmatrix} x'_d i_d \\ x'_q i_q \end{bmatrix} \quad \text{(5)}$$

In (2), vectors $w_k$ and $v_k$ are the process noise and measurement noise, respectively. Their mean and variance are denoted by (6) and (7), respectively. In (6) and (7), symbol $E(*)$ represents the expected value; $Q_k$ and $R_k$ denote the covariance matrixes of the process noise and the measurement noise, respectively.

$$E(w_k) = 0 \quad \text{(6.a)}$$
$$E(v_k) = 0 \quad \text{(6.b)}$$
$$E(w_k w_k^T) = Q_k \quad \text{(7.a)}$$
$$E(v_k v_k^T) = R_k \quad \text{(7.b)}$$

## III. PROPOSED NON-LINEARITY INDEXES

This section discusses two non-linearity indexes, which is proposed to quantify the nonlinearity levels of the state transition function and measurement function. The 1[st] order Taylor expansion can be calculated by ignoring the higher order of Taylor series as shown in (8) and (9). Here, the symbol $\delta x$ which is a perturbation to state $x_k$, can be calculated as the state change between two consecutive time steps as in (10).

$$\Phi(x_k + \delta x) \approx \Phi(x_k) + \left.\dfrac{\partial \Phi(x)}{\partial x}\right|_{x_k} \delta x + \text{higher order terms} \quad \text{(8)}$$

$$h(x_k + \delta x) \approx h(x_k) + \left.\dfrac{\partial h(x)}{\partial x}\right|_{x_k} \delta x + \text{higher order terms} \quad \text{(9)}$$

$$\delta x = x_{k+1} - x_k \quad \text{(10)}$$

The EKF approximates the non-linear function using 1[st] order Taylor expansion. In highly non-linear systems, the linearization error introduced by ignoring the higher order can be significant. Therefore, the EKF can provide an optimal estimation accuracy only when both the state and measurement functions are linear or quasi-linear [24]. For a system to be deemed linear or quasi-linear, the linearization errors should be insignificant as they are compared with the variance of process and measurement noise.

To quantify the non-linearity levels of the state transition function and measurement function, two non-linearity indexes of $n(\Phi)$ and $n(h)$ are defined by (11) and (12). Here, symbols $\varepsilon_\Phi$ and $\varepsilon_h$ in (11) and (12) are the difference between the perturbed state or measurement from their corresponding linear approximation. Their magnitudes depend on both the shapes of $\Phi(x)$ and $h(x)$ and the perturbation of the states (i.e., $\delta x$). If $\Phi(x)$ and $h(x)$ are linear or quasi-linear, $\delta x$ is close to 0. Therefore, $\varepsilon_\Phi$ and $\varepsilon_h$ shall approach 0. Otherwise, $\varepsilon_\Phi$ and $\varepsilon_h$ should be significantly different from 0. To evaluate the influence of the non-linearity on the prediction, $\varepsilon_\Phi$ and $\varepsilon_h$ are normalized by the covariance of the process noise (i.e., $Q_k$) and the covariance of the measurement noise (i.e., $R_k$) in (11.b) and (12.b), respectively. With this normalization, $n(\Phi)$ and $n(h)$ shall be non-negative scalars. Therefore, when the linearization errors are much less than corresponding noise, $n(\Phi) \ll 1$ and $n(h) \ll 1$, the state function and measurement function can be considered quasi-linear. Otherwise, the linearization errors are significant in comparison with the noises. A model can be deemed quasi-linear when both $n(\Phi)$ and $n(h)$ are considerably less than 1.

**Nonlinearity index $n(\Phi)$ for state transition function:**

$$\underbrace{\varepsilon_\Phi = \Phi(x_k + \delta x) - \left[\Phi(x_k) + \frac{\partial \Phi(x)}{\partial x}\bigg|_{x_k} \delta x\right]}_{\text{approximation error}} \quad (11.a)$$

$$n(\Phi) = \varepsilon_\Phi^T Q_k^{-1} \varepsilon_\Phi \quad (11.b)$$

**Nonlinearity index $n(h)$ for measurement function:**

$$\underbrace{\varepsilon_h = h(x_k + \delta x) - \left[h(x_k) + \frac{\partial h(x)}{\partial x}\bigg|_{x_k} \delta x\right]}_{\text{approximation error}} \quad (12.a)$$

$$n(h) = \varepsilon_h^T R_k^{-1} \varepsilon_h \quad (12.b)$$

When the system non-linearity is severe, multi-step prediction can be repeated a few times to reduce the perturbation of the state, i.e., $\delta x$, in (10), (11) and (12), which will reduce the negative impact of non-linearity on the EKF's estimation accuracy.

## IV. EKF AND AMSP

In the practical application of the EKF to estimate power system dynamic states, its prediction step and correction step may be performed with different time cycles for model simulation and measurement update. For power system dynamic simulation, a few millisecond (e.g. 5 ms) can capture the nonlinearity, while the PMU's typical sampling rates are 30 and 60 samples/s. Consequently, the measurement update phase of the EKF will require more computational capacity than model simulation.

The different phases of the EKF can be iterated at different rates. For example, covariance matrix propagation and update have highest computational complexity [24]. Therefore, the Kalman filter equation involving the covariance matrix, P, imposes high computational load. Accordingly, state vector propagation is iterated at higher rates than the covariance propagation based on Fig. 1.

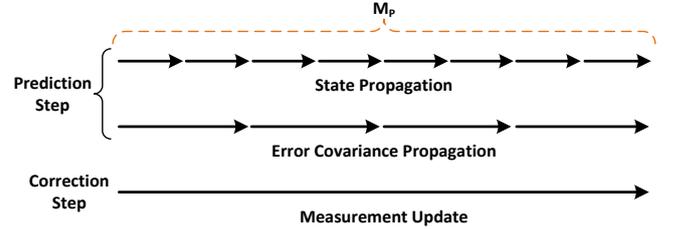

Fig. 1. Kalman filter iteration rate

Based on the determined non-linearity indexes in the previous section, this section proposes an **AMSP** approach to modify the EKF by adaptively adjusting the multi-step prediction index ('$M_p$') to achieve a well-informed trade-off between computation time and estimation accuracy. The modified 3-step EKF procedure is illustrated by Fig. 2 and detailed as follows:

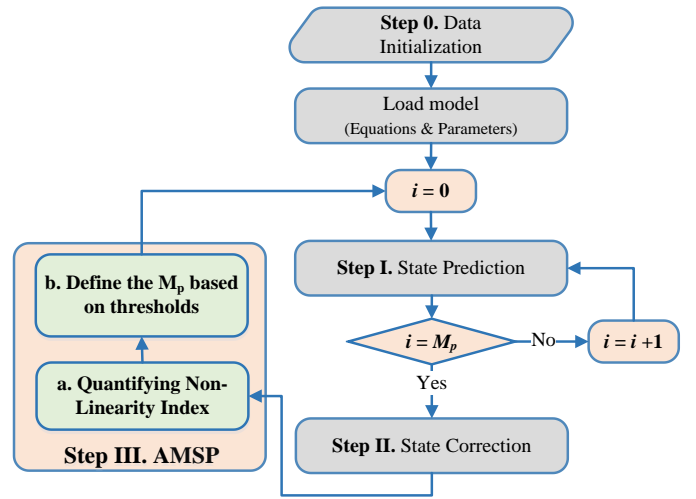

Fig. 2. Modified EKF using the **AMSP** approach

**Step 0 - Initialization:**

$$\begin{cases} \hat{x}_0(+) = E(x_0) & (13.a) \\ P_0(+) = E\left[(x_0 - \hat{x}_0(+))(x_0 - \hat{x}_0(+))^T\right] & (13.b) \end{cases}$$

**Step I - Prediction:**

By applying the multi-step predication to tradition EKF, the prediction step repeats for a few time cycles before performing the correction step. Equations (14) and (15) show the filter with two prediction steps. The Euler method is implemented in this example.

*Prediction step 1:*

$$\hat{x}_{k-1/2}(-) = \hat{x}_{k-1}(+) + \Phi(\hat{x}_{k-1}(+))\frac{\Delta t}{2} \quad (14.a)$$

$$\Phi_{k-1}^{[1]} \approx \left.\frac{\partial \Phi(x)}{\partial x}\right|_{x=\hat{x}_{k-1}(-)} \quad (14.b)$$

$$P_{k-1/2}(-) = \Phi_{k-1}^{[1]} P_{k-1}(+) \Phi_{k-1}^{[1]T} + Q_{k-1} \quad (14.c)$$

*Prediction step 2:*

$$\hat{x}_k(-) = \hat{x}_{k-1/2}(+) + \Phi(\hat{x}_{k-1/2}(+))\frac{\Delta t}{2} \quad (15.a)$$

$$\Phi_{k-1/2}^{[1]} \approx \left.\frac{\partial \Phi(x)}{\partial x}\right|_{x=\hat{x}_{k-1/2}(-)} \quad (15.b)$$

$$P_k(-) = \Phi_{k-1/2}^{[1]} P_{k-1/2}(+) \Phi_{k-1/2}^{[1]T} + Q_{k-1/2} \quad (15.c)$$

**Step II - Correction:**

$$\hat{x}_k(+) = \hat{x}_k(-) + \bar{K}_k \left[z_k - h_k(\hat{x}_k(-))\right] \quad (16.a)$$

$$H_k^{[1]} = \left.\frac{\partial h(x)}{\partial x}\right|_{\hat{x}_k(-)} \quad (16.b)$$

$$\bar{K}_k = P_k(-) H_k^{[1]T} \left[ H_k^{[1]} P_k(-) H_k^{[1]T} + R_k \right]^{-1} \quad (16.c)$$

$$P_k(+) = \left\{ I - \bar{K}_k H_k^{[1]} \right\} P_k(-) \quad (16.d)$$

where the "(+)" indicates that the estimate is *a posteriori*, and $P$ is the state covariance matrix. $\Phi_{k-1}^{[1]}$ is the Jacobian matrix of the state transition matrix at step $k$-1 and $H_k^{[1]}$ is the Jacobian matrix of the measurement function at step $k$.

**Step (III) – AMSP:**

To make a trade-off between computation time and estimation accuracy, the **AMSP** approach is proposed to define the $M_p$ based on non-linearity indexes. The **AMSP** approach consists of two sub-steps. *a)* the non-linearity indexes of the state transition function and the measurement function [i.e., $n(\Phi)$ and $n(h)$] are calculated using (11) and (12). And *b)* the multi-step predication factor '$M_p$' is determined based on the non-linearity indexes and predefined thresholds. This step has three parameters: multi-step predication factor ('$M_{p,k}$'), upper threshold ($U$) and lower threshold ($L$). The nonlinearity indexes at each step are compared with thresholds $U$ and $L$ to determine the '$M_p$'. If one of the $n(\Phi)$ or $n(h)$ is greater than $U$, '$M_p$' will be increased to reduce the linearization errors. On the other hand, if both $n(\Phi)$ and $n(h)$ are less than the $L$, '$M_p$' will be decreased to reduce the computational complexity. The proposed **AMSP** is tugged into the EKF to adaptively adjust the '$M_p$' based on the levels of the system non-linearity. For larger $n(\Phi)$ or $n(h)$, a larger '$M_p$' is selected to reduce the non-linearity impact on DSE accuracy. For smaller $n(\Phi)$ and $n(h)$, a smaller '$M_p$' is assigned to reduce the computation time. The number of the repeats of prediction step is calculated by $2^{M_p}$. For a linear system, both $n(\Phi)$ and $n(h)$ are zero, hence $M_p = 0$ and number of predication step repeats is one.

## V. STUDY APPROACHES

Performance of the proposed **AMSP** approach is evaluated using the two-area four-machine system [1] (Fig. 3) with model stored in "d2asbeghp.m" to generate the simulated PMU data. The simulation is performed using the Power System Toolbox (PST) [29]. To mimic real system responses, a three-phase fault is applied to the sending end of the line between buses 100 and 200. Two sets of data are generated as follows:

1) *Lightly damped data* are generated to evaluate the estimation accuracy during severe oscillation responses. To incur severe oscillations, all the power system stabilizers (PSS) are turned off. The simulation is performed for 30 s. The fault is applied at 10.1 s.

2) *Well damped data* are generated to evaluate the balance between computation time and estimation accuracy under a normal operation condition. All PSSs are turned on. As a result, the transient responses to the fault are damped very quickly. The simulation is performed for 720 s (i.e. 12 min). The fault is applied at 60.1 s.

The fault is cleared in 50 ms after the fault at the near end (bus 100), and in 100 ms after the fault at the remote end (bus 200). To reduce integration errors and capture the dynamics, the simulation time step is chosen to be 0.001 s.

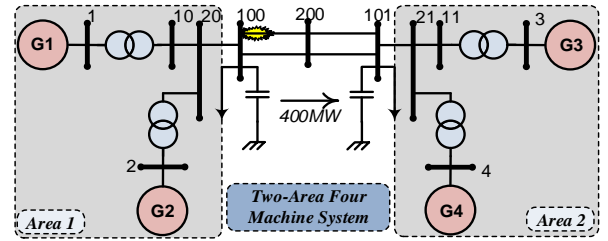

Fig. 3. The two-area four-machine system [29]

Assume that bus 1 has a PMU, which measures the voltage phasor and current phasors, i.e., $e_I$, $e_R$, $i_I$ and $i_R$ in (3). To mimic the field measurements from the PMUs, the simulation data is decimated to 25 samples/s and measurement noise is added. Considering noise introduced by PT and CT, 4.0% of noise in total vector errors (TVE) [24] are added to the current and voltage phasors. Also, 4.0% of noise is added to the $E_{fd}$ and $T_m$.

To initialize the EKF, covariance $P_0$ is set to be 10 times of the largest changes of the states to reflect the uncertainty of the initial states. The output variance $R_k$ is set to be 4.0% of measurements i.e., $diag([0.04, 0.04])^2$. $Q_k$ is set to be $Q_k = 0.04 * \max(\text{abs}(\text{diff}(x_{1:N})))$ which corresponds to the 4.0% of the largest state changes in each step.

Because of the randomness of the process and measurement noises, the Monte-Carlo (MC) method is used to evaluate the statistical performance of the algorithms. The MC method generates various instances of random noises by randomly sampling noise's distribution [17]. Based on the MC simulation, the mean square error (MSE) is defined in (17) to quantify the estimation errors at the $k^{th}$ time-step. The mMSE is defined in (18) to quantify the overall estimation errors.

$$MSE(\hat{x}_k) \overset{\Delta}{=} \frac{1}{N}\sum_{n=1}^{N}(\hat{x}_{k,n} - x_{k,True})^2 \qquad (17)$$

$$mMSE = \frac{1}{K}\sum_{k=1}^{K} MSE(\hat{x}_k) \qquad (18)$$

Here, $\hat{x}_{k,n}$ is the estimated states in the $n^{th}$ Monte-Carlo test case, while $x_{k,True}$ is the corresponding true state. The symbol $K$ is the length of the measurement data, while $N$ represents the total number of Monte-Carlo trials. "Note that $N$ should be large enough to make estimated MSEs converge and small enough to avoid unnecessary computational burden" [17]. This paper uses $N = 100$.

For a real-time application, computation time is another important factor of an algorithm. To estimate states in real time, estimation of the current time step must be finished before the measurements of the next time step arrive. In other words, a state estimation algorithm must be fast enough to keep up with the measurement data flow. A filtering algorithm is effective if it uses less computation time to compute the state estimation with high accuracy.

To count the computation time, all the simulations are implemented using MATLAB® version *R2013b* and tested on a computer with a 3.5-GHz processor, 16 GB of memory, and a 64-bit operating system. The computation timer is started right before the prediction step of the first measurement and ended right after the correction step of the last measurement. For the Monte-Carlo simulation, the mean value of the computation time is calculated and used for evaluating performance.

## VI. CASE STUDIES

This section evaluates the performance of the proposed **AMSP** approach through simulation. The estimation accuracy (in term of mMSE) and computation time of the EKF are compared in following scenarios:

(A) the impact of multi-step predication,

(B) the performance of the **AMSP** approach.

### A. Impact of Multi-step Prediction

The goal of this scenario is to evaluate the impact of multi-step prediction [22] on the EKF's estimation accuracy and nonlinearity indexes. Using the *lightly damped data*, the dynamic states of generator G1 are estimated using the EKF for 100 sets of MC simulation. The representative estimation results dynamic states is summarized in Fig. 4.

In Fig. 4, it can be observed that estimated states for the 100 sets of data using both the standard EKF ($M_p = 0$) and the constant multi-step prediction (CMSP) approach ($M_p = 3$) converge. The estimation errors for the case with multi-step

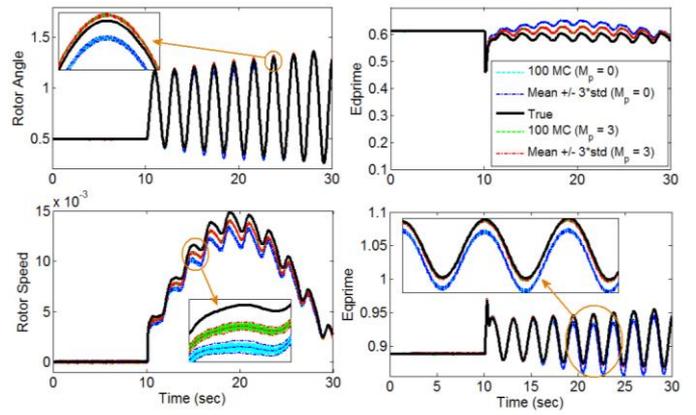

Fig. 4. Estimated states of G1 with $M_p = 0$ and $M_p = 3$ using 100 sets of MC simulations for the *lightly damped data*.

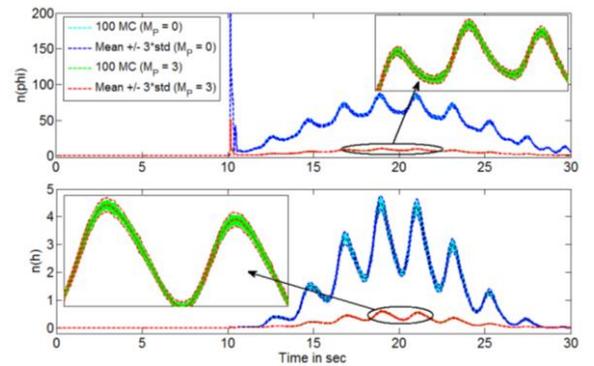

Fig. 5. Nonlinearity indexes of the state transition function and measurement function for $M_p = 0$ and $M_p = 3$ using 100 sets of MC simulations for the *lightly damped data*.

prediction is smaller than the CEKF approach during the transient responses after 10.1 s.

Fig. 5 summarizes the corresponding nonlinear indexes for the standard EKF and CMSP. From Fig. 5, it can be observed that the nonlinearity indexes during the transient responses after 10 s are larger than those during the steady-state responses (before 10 s). The observation indicates that the transient responses will incur high nonlinearity. Also, it can be observed that by repeating prediction step a few time, for example for $M_p = 3$ is 8 times ($2^3 = 8$), the nonlinearity indexes decrease (as shown in Fig. 5) and the estimation errors also decrease (as shown in Fig. 4).

### B. Adaptive Multi-Step Predication (**AMSP**) Approach

The system non-linearity levels change with operation conditions. When non-linearity levels change, a constant *'$M_p$'* used in the CSMP may not adequately capture the non-linearity. To avoid such drawbacks, this paper proposes the **AMSP**, which adaptively adjusts *'$M_p$'* according to the levels of non-linearity. This subsection evaluates the performance of the proposed **AMSP** in comparison to the CMSP. To make a well-informed trade-off between estimation accuracy and computation time, it is important to set proper thresholds, i.e., $U$ and $L$, for the nonlinearity indexes. In this paper, $U = 0.3$ and $L = 0.005$ are used. Note that the nonlinearity indexes are the relative linearization errors with respect to the model noise.

Therefore, $U = 0.3$ indicates that when the linearization errors are greater than 30% of model noise, '$M_p$' shall be increased to repeat more prediction step and reduce the linearization errors. Similarly, $L_i = 0.005$ indicates that when the linearization errors are less than 0.5% of model noise, '$M_p$' shall be reduced to decrease the computation time. Users may choose different $U$ and $L$ according to the needs of their particular applications.

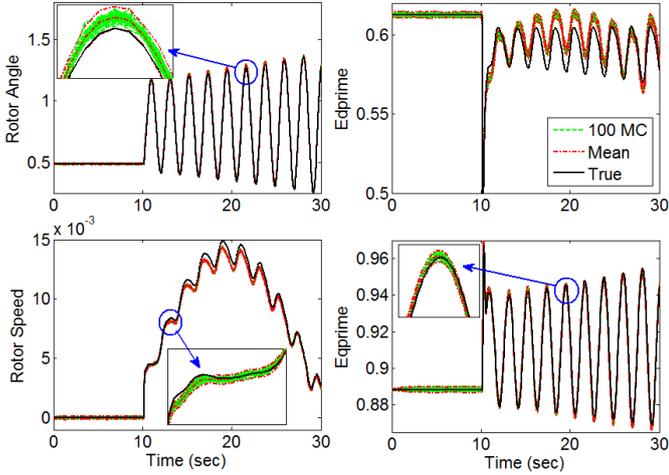

Fig. 6. Estimated states of G1 in the two-area model with the **AMSP** approach using 100 sets of MC simulations for the *lightly damped data*.

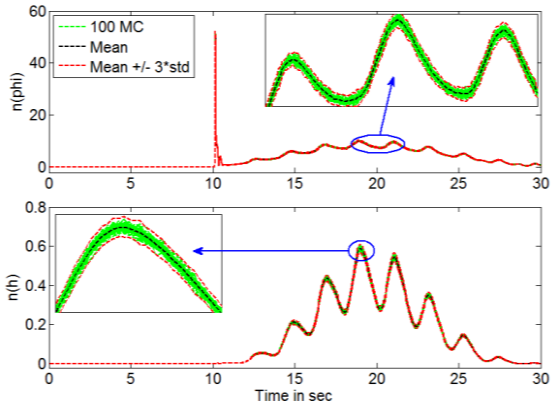

Fig. 7. Nonlinearity indexes of the state transition function and measurement function of G1 with $M_p = 5$ for the *lightly damped data*.

For the *lightly damped data* the estimated states, nonlinearity indexes and corresponding $M_p$ changes at each time step using 100 sets of MC using the **AMSP** approach are summarized in Fig. 6, Fig. 7 and Fig. 8. From Fig. 6 it is clear that the fault is applied at 10.1 s. During the steady-state response, the corresponding nonlinearity indexes (Fig. 7) are relatively small (close to 0). Therefore $M_p$ is set 0, to decrease the computation time (Fig. 8). On the other hand, during the transient responses, the nonlinearity indexes increase. Based on that, the corresponding $M_p$ is increased to reduce the negative impact of nonlinearity on the estimation accuracy. The $M_p$ changes at each step of time based on nonlinearity indexes is shown at Fig. 8.

The mMSE and computation time of the **AMSP** approach is compared with the CMSP ($M_p = 5$) for the *lightly damped data* with 100 set of MC and summarized in TABLE I. To facilitate the comparison, the 30 s data are divided into three 10 s segment. From TABLE I. it can be observed that the proposed **AMSP** approach uses much shorter computation time than the CMSP approach to achieve the similar mMSE. For example, at Seg#1, with same estimation accuracy, the proposed approach decrease the computation time by around 95%.

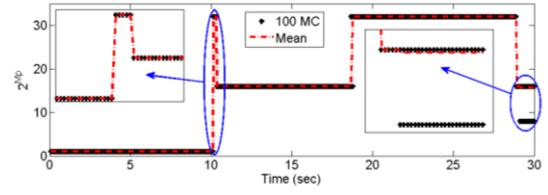

Fig. 8. $M_p$ changes based on the proposed **AMSP** approach for G1 with 100 MC for the *lightly damped data*.

TABLE I. COMPARISON OF THE **AMSP** AND CMSP ($M_P = 5$)

| States (mMSE) | | Whole | Seg #1 | Seg #2 | Seg #3 |
|---|---|---|---|---|---|
| $\delta$ | AMSP | 1.85E-04 | 3.93E-06 | 2.95E-04 | 2.64E-04 |
| | CMSP | 1.91E-04 | 3.93E-06 | 3.01E-03 | 2.69E-06 |
| $\Delta\omega$ | AMSP | 2.85E-07 | 1.78E-09 | 4.59E-07 | 3.96E-07 |
| | CMSP | 2.89E-07 | 1.79E-09 | 5.58E-07 | 3.97E-07 |
| $e'_d$ | AMSP | 2.82E-04 | 3.61E-07 | 4.68E-04 | 3.8E-04 |
| | CMSP | 2.83E-04 | 4.21E-07 | 4.72E-04 | 3.81E-04 |
| $e'_q$ | AMSP | 3.81E-06 | 2.25E-07 | 5.26E-06 | 5.96E-06 |
| | CMSP | 3.79E-06 | 2.28E-07 | 5.24E-06 | 6.03E-06 |
| Computation Time (s) | AMSP | 26.91 | 1.07 | 10.02 | 15.83 |
| | CMSP | 52.73 | 17.71 | 17.50 | 17.51 |

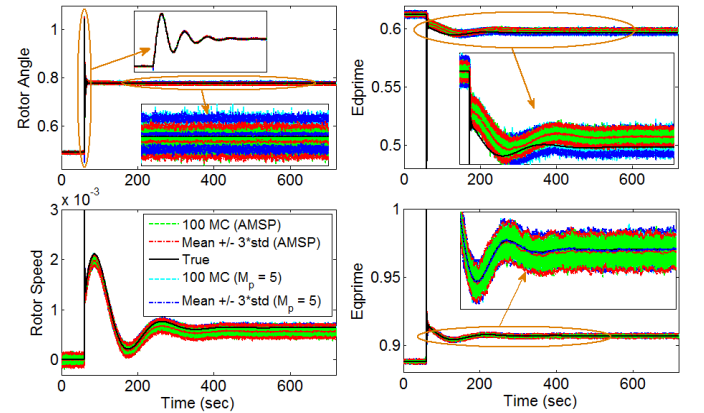

Fig. 9. Estimated states of G1 by the CMSP ($M_p = 5$) and the **AMSP** approaches using 100 sets of MC simulations for the *well damped data*.

For the *well damped data*, the estimated states of generator G1 using the **AMSP** and the CMSP approaches are compared in Fig. 9. The nonlinearity index for the CMSP ($M_p = 5$) is shown in Fig. 10. And the **AMSP** corresponding $M_p$ is summarized in Fig. 11. It can be observed from Fig. 11 that the **AMSP** approach changes the values of $M_p$ based on the nonlinearity indexes. For example, when the nonlinearity indexes are small during the steady-state responses (Fig. 10), the **AMSP** approach uses $M_p = 0$ at the prediction step (Fig. 11). On the other hand, when the nonlinearity is severe during the transient responses, the **AMSP** approach increases $M_p$ at the prediction step to reduce the negative impact of the nonlinearity on the estimation accuracy. The figures suggested that based on the nonlinearity indexes, the **AMSP** approach uses different $M_p$ to get a trade-off between computation time and estimation accuracy. In contrast, the CMSP approach has $M_p = 5$ all the time.

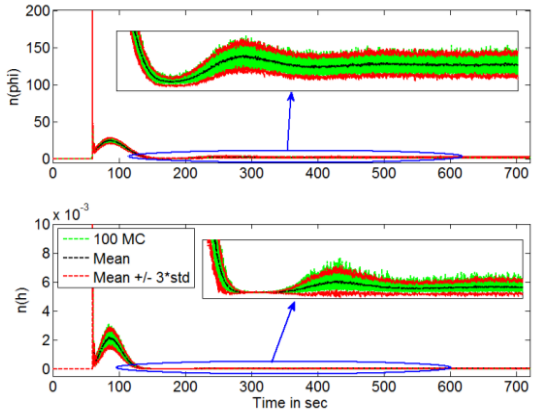

Fig. 10. Nonlinearity indexes of the state transition function and measurement function of G1 with $M_p = 5$ for the *well damped data*.

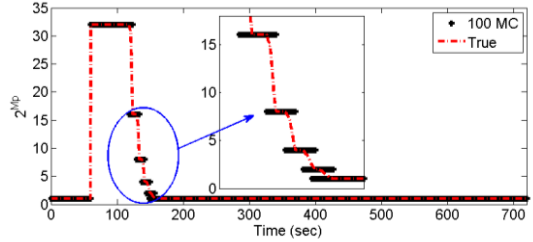

Fig. 11. $M_p$ changes based on the proposed **AMSP** approach for $G_1$ with 100 MC for the *well damped data*.

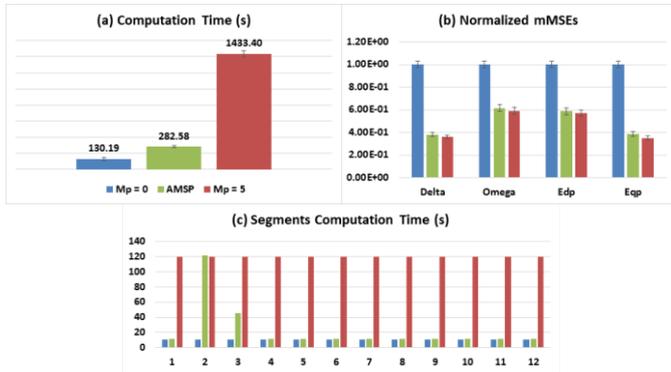

Fig. 12. Comparison of computation time and mMSEs of the EKF ($M_p$=0), **AMSP** and CMSP ($M_p = 5$) using the *well damped data.*

To reveal the overall performance of the **AMSP** approach, comparison of computation time and estimation accuracy is made among three approaches: EKF ($M_p = 0$), **AMSP** and CMSP ($M_p = 5$). The resulting computation time and estimation accuracy are summarized in Fig. 12. Fig. 12 (a) shows that the computation time of the EKF ($M_p = 0$) approach is shortest while the computation time of the CMSP ($M_p = 5$) approach is longest. Fig. 12 (b) shows that the CEKF approach has the highest normalized mMSEs. On the other hand, the mMSEs of the **AMSP** and the CMSP approaches are similar. Note that the **AMSP** has almost same mMSEs as the CMSP approach while the computation time of the **AMSP** is much shorter. To facilitate comparison, the *well damped data* (720 s) are divided to twelve 60-s segments for studying the computation time in Fig. 12 (c). Note that segment 1 and segments 4–12 are mainly steady-state responses. As a result, the **AMSP** approach has 1 step prediction (Fig. 11). As such, its corresponding computation time is considerably shorter than the CMSP. It can be concluded that the **AMSP** approach uses much shorter computation time than the CMSP approach to achieve the similar mMSEs.

## VII. CONCLUSIONS

This paper proposes an **AMSP** approach to achieve good trade-off between estimation accuracy and computation time of the EKF in dynamic state estimation of a synchronous machine. Two indexes are proposed to quantify the nonlinearity levels of the state transition function and measurement function. When the nonlinearity levels are high, Kalman filter's prediction step is repeated a few time cycles to decrease its time step, which reduces the negative impact of non-linearity on state estimation accuracy. When the nonlinearity levels are low, the prediction step is repeated less (or sometimes none) to shorten computation time.

Using the two-area four-machine, it is shown that the nonlinearity levels are high during the transient responses and low during steady-state responses. The proposed **AMSP** approach can detect the changes in nonlinearity levels and adaptively adjust the prediction step factor '$M_p$' to achieve a desired trade-off between estimation accuracy and computation time. Such a trade-off is critical for removing the barriers to adopting such estimation technologies for real-time applications.

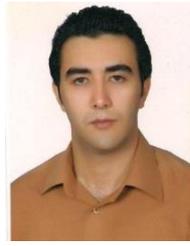

**Shahrokh Akhlaghi** (S'14) received the M.S. degree in electrical engineering from the Amirkabir University of Technology, Tafresh, Iran, in 2010. He is currently working toward the Ph.D. degree in electrical engineering from Binghamton University, State University of New York (SUNY), Binghamton, NY, USA. He was a Lecturer in Department of Electrical and Computer Engineering at Tehran North Branch of Islamic Azad University, Tehran, Iran from 2010 to 2013.

His research interests include power system dynamics, state estimation, phasor measurement unit, Kalman filtering, distributed generation and islanding detection of DGs. Shahrokh Akhalghi is the chairman of the IEEE PES Student Branch Chapter and IEEE Young Professional affinity group at Binghamton.

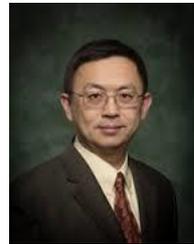

**Ning Zhou** (S'01–M'05–SM'08) received the Ph.D. degree in electrical engineering with a minor in statistics from the University of Wyoming, Laramie, WY, USA, in 2005. He is an Assistant Professor at the Electrical and Computer Engineering Department of the Binghamton University, Binghamton, NY, USA. He was with Pacific Northwest National Laboratory (PNNL) as a power system engineer from 2005 to 2013.

His research interests include power system dynamics and statistical signal processing. Dr. Zhou is a senior member of the IEEE Power and Energy Society (PES).